# A Compact Single-Flange Low-Energy Inverse Photoemission Spectrometer with Elliptical-Mirror Photon Collection

Jong-Am Hong [a,1], Min-Jae Maeng [a,1], Kyu-Myung Lee [a], Seunggu Lee [a], In-Sung Park [a], Yongsup Park [a,b,*]

[a] *Department of Physics and Research Institute for Basic Sciences, Kyung Hee University, 26 Kyungheedae-ro, Dongdaemun-gu, Seoul 02447, Republic of Korea*

[b] *Department of Information Display, Kyung Hee University, 26 Kyungheedae-ro, Dongdaemun-gu, Seoul 02447, Republic of Korea*

*** Corresponding author**

*E-mail address*: parky@khu.ac.kr (Y. Park)

[1] These authors contributed equally to this work.

## Abstract

Low-energy inverse photoemission spectroscopy (LEIPS) is limited by weak photon emission, so photon collection optics largely determine its sensitivity. We compare four photon-collection configurations, each addressing a limitation of its predecessor: a detector beside the sample without collection optics (no-mirror), an off-axis parabolic (OAP) mirror delivering a collimated beam to a head-on photomultiplier tube (PMT), external optics coupling the OAP output to a side-on PMT with about twice the quantum efficiency (OAP-external), and an in-vacuum elliptical mirror focusing the emitted light onto the detector (elliptical). Ray-tracing shows that the elliptical mirror delivers 91.9% of the emitted photons to the detector, compared with 63.3% for the OAP mirror and 3.06% without a mirror, while a plane mirror folding the light path by about 90° allows the electron gun, collection optics and detector to fit on a single 8-inch flange. For $C_{60}$ films, the elliptical configuration gives the highest lowest unoccupied molecular orbital (LUMO) intensity of the four, 271 counts per second per μA of sample current, 1.74 and 3.23 times those of the OAP and OAP-external configurations and about 45 times that of the no-mirror configuration. The gain over the OAP configuration is comparable to the 1.45-fold increase predicted by ray-tracing, but the filter and detector also differed, so the geometric contribution could not be isolated. The OAP-external configuration gave only 0.54 times the OAP signal despite its higher quantum efficiency. High sensitivity therefore requires both a large collection solid angle and a short optical path kept inside the vacuum.

*Keywords*: Low-energy inverse photoemission spectroscopy (LEIPS), Unoccupied states, Elliptical mirror, Photon collection efficiency, Instrument design

## 1. Introduction

Energy level alignment at organic semiconductor interfaces sets the carrier injection barriers and governs device operation, so describing this alignment requires both the occupied and the unoccupied levels [1,2]. Photoemission spectroscopy (PES) measures the occupied levels directly, whereas the unoccupied levels are often derived from the occupied levels by adding the measured optical gap and an estimated exciton binding energy. The exciton binding energy is difficult to measure directly, so the resulting unoccupied levels are uncertain [3].

Inverse photoemission spectroscopy (IPES) measures the unoccupied levels directly and was introduced in the vacuum ultraviolet (VUV) range by Dose [4]. It has investigated the unoccupied electronic structure of metals and semiconductors [5] and more recently of organic thin films [6]. Its sensitivity is extremely low because the cross section for an incoming electron to emit a photon is several orders of magnitude smaller than that of photoemission [7]. The cross section is intrinsic to the transition and a high electron current can damage samples, so the signal has to be raised by collecting and detecting more of the emitted photons. Early momentum-resolved isochromat spectroscopy did so by placing the sample and the detector at the two foci of an elliptical geometry [8], and other optics have since been used to cover a large solid angle [9].

Low-energy inverse photoemission spectroscopy (LEIPS) uses low-energy electrons to reduce damage to organic samples [10,11] and has become a standard method for measuring the electron affinity of organic solids [12]. Conventional IPES detects photons near 10 eV with bandpass counters whose entrance window and fill gas define the detection band [13–15]. LEIPS instead detects photons near 5 eV (250 nm) either with a grating spectrometer [16] or with a bandpass filter in front of a photomultiplier tube (PMT) [17]. Ordinary transmissive and reflective optics can be used at this detection wavelength, so the collection optics can be designed freely.

We previously compared a detector placed next to the sample (no-mirror setup, **Fig. 1a**) with an off-axis parabolic (OAP) mirror and a head-on PMT brought close to it through a re-entrant vacuum window (OAP setup, **Fig. 1b**) [18]. This OAP mirror increased the lowest unoccupied molecular orbital (LUMO) intensity of a $C_{60}$ film on Ag by a factor of 26 and enabled thickness-dependent energy level alignment measurements of organic films. A side-on PMT offers about twice the quantum efficiency of the head-on PMT but receives light through its side wall, so it is difficult to place behind a re-entrant window. We therefore took the collimated light outside the chamber and focused it onto the PMT with lenses and mirrors (OAP-external setup, **Fig. 1c**). The external setup allowed filter exchange and alignment without breaking vacuum, but its signal fell to 0.54 times that of the OAP setup despite the higher quantum efficiency of the side-on PMT. It also needed separate ports for the electron gun and the detection optics, and its longer optical path contained more elements where light could be lost.

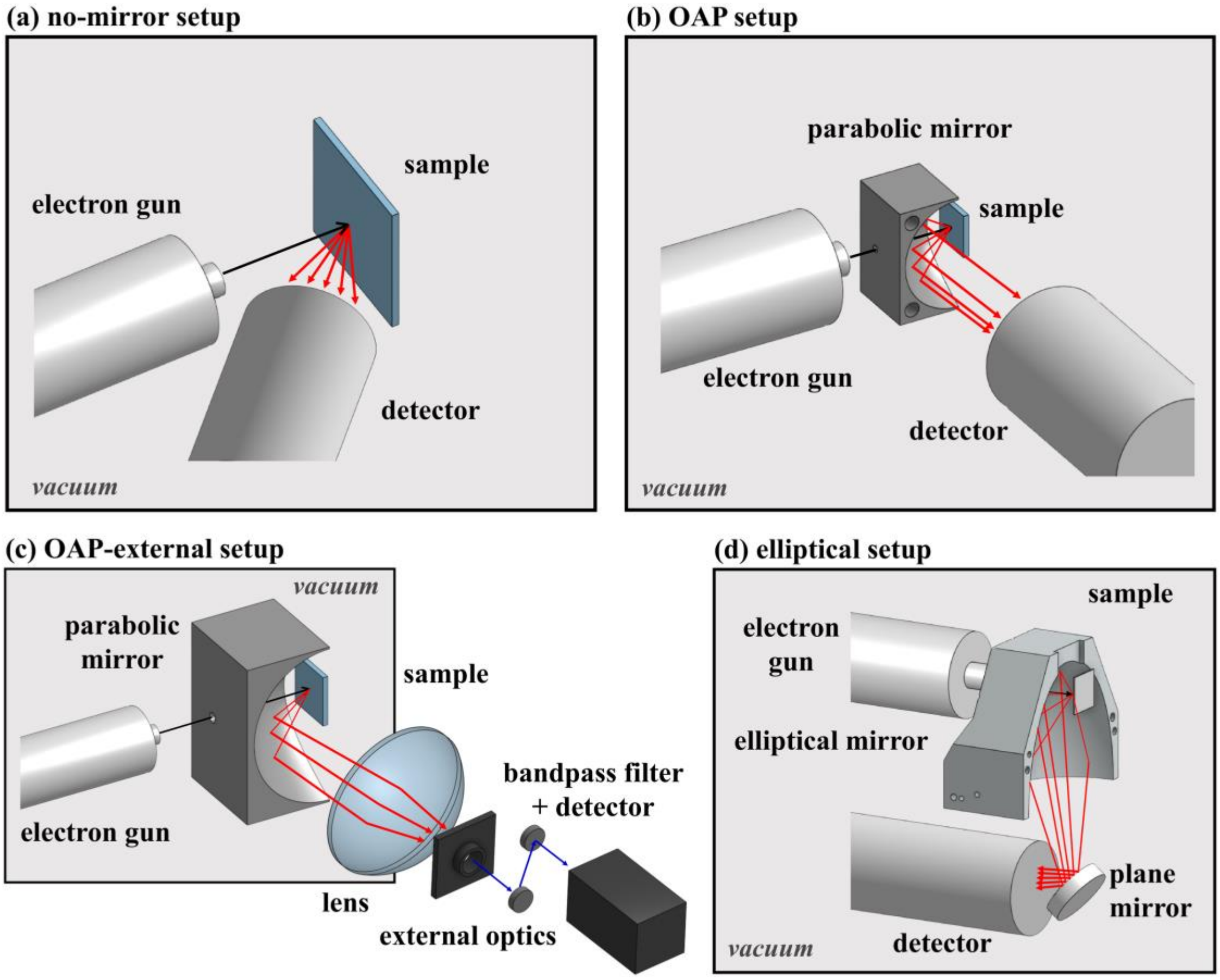

**Figure 1.** Schematics of the four photon-collection setups. (a) No-mirror, (b) off-axis parabolic (OAP), (c) OAP-external and (d) elliptical setups. The black arrows show the electron beam, the red lines the photon paths from the sample, and the blue arrows in (c) the photon paths outside the vacuum chamber. The detector is a head-on photomultiplier tube (PMT) in (a) and (b), a side-on PMT with about twice the quantum efficiency at the end of the external optics in (c) and a photon counting head in (d). In (a), (b) and (d) it sits directly behind a re-entrant vacuum window. The vacuum windows are not drawn.

To increase the collection while keeping the optical path short and inside the vacuum, we used an elliptical mirror that focuses the light from the sample at one focus onto the detector near the other focus without lenses (elliptical setup, **Fig. 1d**). Kashimoto et al. also used an elliptical mirror for angle-resolved LEIPS measurements with a rotatable sample [19]. We placed our mirror very close to the sample for efficient collection, so the incidence angle is fixed and momentum-resolved measurements, which require varying this angle [19–21], are not possible. In return, the optical path from the sample to the second focus is only 110 mm, and a plane mirror folds it by about 90° so that the detector lies parallel to the electron gun, and the electron gun, both mirrors and the detector fit on a single 8-inch flange. Our instrument combines LEIPS and PES in the same chamber, as in the combined system of Babbe *et al.* [22].

Here we compare the sensitivity of the four setups using $C_{60}$/Ag films prepared by the same procedure. Ray-tracing shows that 91.9% of the emitted photons reach the detector window in the elliptical setup. Its LUMO intensity of 271 counts per second per μA of sample current is 1.74 times that of the OAP setup and about 45 times that of the no-mirror setup, the highest of the configurations we have tested so far. Together with the loss in the OAP-external setup, this shows that high sensitivity requires both a large collection solid angle and a short optical path kept inside the vacuum.

## 2. **Methods**

### *2.1. Experimental*

All measurements were performed in situ in an ultra-high vacuum (UHV) chamber with a base pressure of $5 \times 10^{-10}$ Torr. Ultraviolet photoemission spectroscopy (UPS, SES-100 analyzer, SCIENTA Omicron GmbH, with He I discharge lamp) and LEIPS were carried out in the same chamber. The samples were prepared in a connected deposition chamber under UHV without air exposure. Ag substrates were cleaned by $Ar^+$ sputtering (PHI 04-303, 2.3 keV) and $C_{60}$ (Sigma-Aldrich) was thermally evaporated on Ag while the thickness was monitored with a quartz crystal microbalance. The $C_{60}$ films, a common LEIPS benchmark [23], were 10 nm thick, except for the 14 nm film measured with the no-mirror setup, and all samples were prepared by the same procedure. For both UPS and LEIPS, binding energies were referenced daily to the Fermi edge of sputter-cleaned Ag. UPS and LEIPS are measured at different positions of the sample manipulator, which translates and rotates the sample between the analyzer and the LEIPS flange. Because the electron gun, the mirrors and the detector are mounted together on one flange, the sample approaches the mirror only in the LEIPS position and is withdrawn from it for UPS and for sample transfer, so the mirror assembly does not interfere with either. For LEIPS, the sample is moved to the manipulator position at which its surface lies at the first focus of the elliptical mirror, 1–2 mm from the mirror aperture, and the same position is used for every measurement. HOMO and LUMO onsets were determined by linear extrapolation of the leading edge of each peak to the background level, and the transport gap is the energy difference between the two onsets.

The electron gun (STAIB NEK-100-1) was the same in all four setups. The cathode voltage was fixed at 7 V and the sample bias was swept from 7 V to 0 V so the electron energy at the sample ranged from 0 to 7 eV [18]. In the OAP, OAP-external and elliptical setups, the collection mirror is located between the gun and the sample and can absorb or deflect low-energy electrons. A bias between 0 and −2 V was applied to the mirror so that the electrons reach the sample through its aperture. Part of the emission current is intercepted by the mirror, but because all spectra are normalized to the current measured at the sample, the intercepted fraction does not enter the comparison between setups. The beam diameter at the sample, about 2 mm, is the value expected for this gun at these energies; it was not measured directly and was used as the source size in the ray-tracing. The same mirror bias was applied when the Ag Fermi edge was measured, so a rigid shift of the energy scale caused by the bias is removed by the daily calibration. The Ag Fermi edge, which appears at low kinetic energy in LEIPS, has a comparable width in the elliptical setup and in the no-mirror setup (Fig. S3); the bias therefore does not visibly distort the spectrum near zero kinetic energy.

The detector was kept in air behind a vacuum window in all four setups. The no-mirror and OAP setups used a head-on PMT (R6834, Hamamatsu), and the OAP-external setup used a side-on PMT (R7154P, Hamamatsu) with about twice its quantum efficiency at 250 nm. The elliptical setup used a Hamamatsu H11870-09-Y003 photon counting head that integrates a metal-package PMT, a counting circuit and a high-voltage supply set at the factory. The no-mirror, OAP and OAP-external setups used a deep ultraviolet (DUV) bandpass filter (BP-250, Edmund Optics) centered at 250 nm with a full width at half maximum (FWHM) of 9.4 nm (≈190 meV) and a peak transmittance of 58%. The elliptical setup used a bandpass filter (Hg01-254, Semrock) centered at 251 nm with an FWHM of 15.7 nm (≈300 meV) and a peak transmittance of 78%. The measured transmittance spectra are shown in **Fig. S1**. The KrF mirrors (Korea Electro-Optics) have a reflectivity above 99.0% at 250 nm. The no-mirror and OAP setups used the bandpass filter alone, whereas the OAP-external and elliptical setups added two KrF mirrors (together with focusing lenses) and one KrF mirror, respectively.

All LEIPS spectra were acquired over the same 7 eV sweep. The no-mirror setup used 0.05 eV steps with a 1 s dwell time and 160 sweeps, which took 22,400 s. The OAP and OAP-external setups used 0.025 eV steps with a 1 s dwell time and 20 sweeps while the elliptical setup used 0.05 eV steps with a 200 ms dwell time and 200 sweeps, giving the same total acquisition time of 5,600 s for these three setups. Spectra were normalized by the sample current and are given in counts per second per μA of sample current (CPS/μA). The sample currents were 2.97, 1.75 and 0.49 μA for the no-mirror, OAP and elliptical setups (Table 1), corresponding to current densities of about 0.95, 0.56 and 0.16 μA/mm$^2$ for the 2 mm beam spot. The lower current in the elliptical setup was a deliberate choice to reduce the charge dose, made possible by its higher count rate.

### *2.2. Optical simulation*

Ray-tracing was performed with Zemax OpticStudio 22.1.2. A total of $10^6$ photons were emitted with a Lambertian angular distribution from a circle of 2 mm diameter on the sample surface, which represents the electron beam spot because the electron trajectories are not part of the ray-tracing. The fraction reaching the effective area of each detector at the vacuum window was counted. Photons striking any surface other than a mirror were treated as absorbed and the mirrors were assumed to reflect all incident light. Window and filter transmittance, mirror losses and quantum efficiency were not included. The same procedure gave 3.06% and 63.3% of the emitted photons for the no-mirror and OAP setups, respectively [18]. The external optical path of the OAP-external setup, from the vacuum window to the side-on PMT, was not ray-traced, so the 63.3% of the OAP setup refers to the light leaving the chamber and not to the light reaching the side-on photocathode.

## 3. **Results and discussion**

### *3.1 Elliptical-mirror optics*

To maximize the collection, the sample surface is placed at the first focus of the elliptical mirror so that the mirror covers nearly the whole half-space above it. A plane mirror just before the second focus then turns the converging light by about 90° toward the detector behind the re-entrant window. In practice, the KrF plane mirror and the detector cannot be placed freely because of the flange and the other parts in the chamber, so we optimized the geometry by ray-tracing together with the mechanical constraints of a three-dimensional model of the assembly.

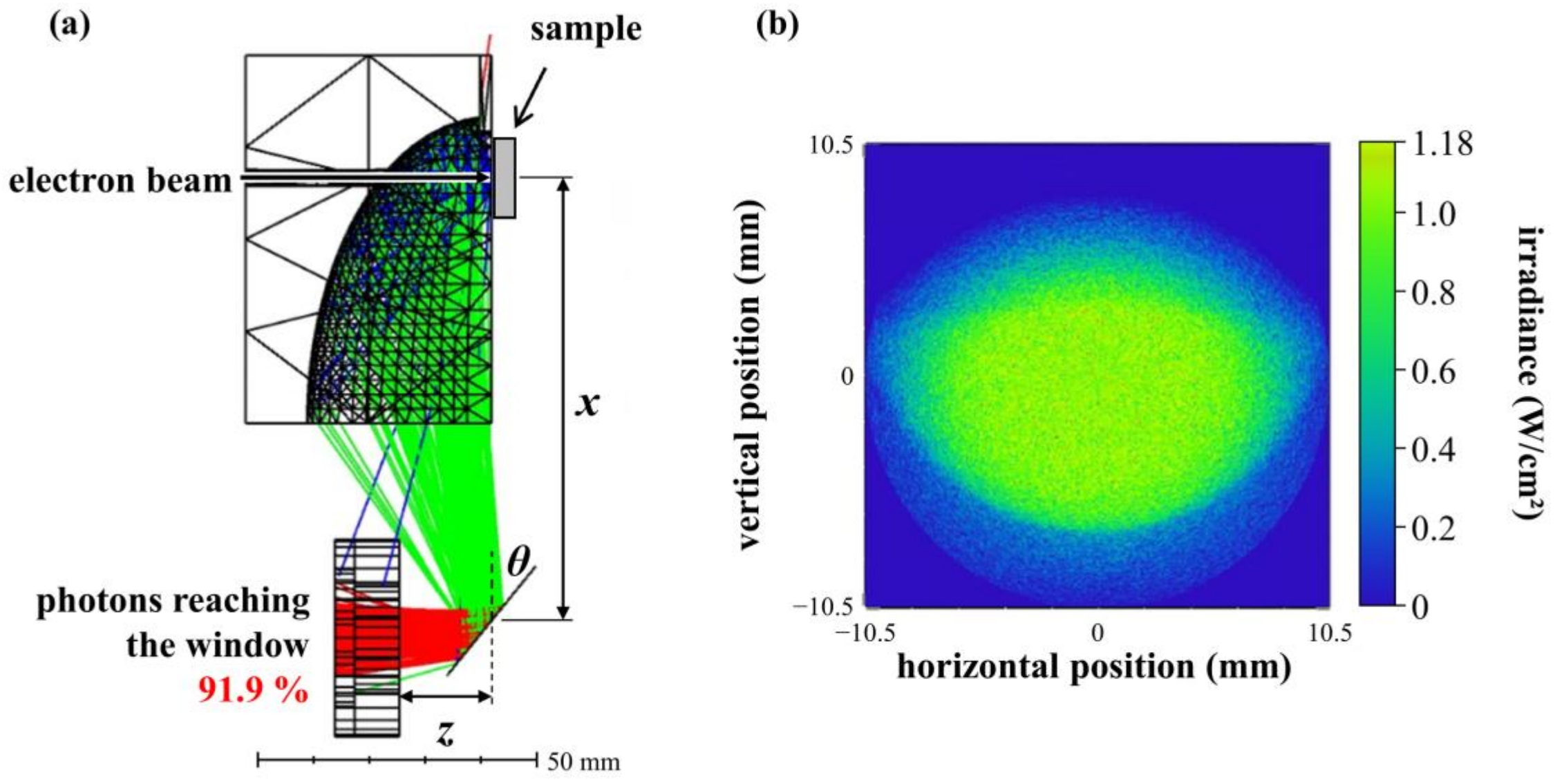

**Figure 2.** Ray-tracing of the elliptical setup. (a) The black arrow is the electron beam, blue the light from the sample to the elliptical mirror, green the light after the elliptical mirror, and red the light after the plane mirror toward the window. The distance $x$ from the sample to the plane mirror, $\theta$ the tilt of the plane mirror, and $z$ the distance from the plane mirror to the window. (b) Distribution of the photons in the plane of the re-entrant window. The axis range of ±10.5 mm is the φ21 mm effective area on which the photons were counted.

The distance $x$ from the sample to the plane mirror, the tilt $\theta$ of the plane mirror and the distance $z$ from the plane mirror to the window define the geometry (**Fig. 2a**), and the size of the elliptical mirror followed from them. The two foci of the ellipse are $2c = 88$ mm apart, but the electron gun and the detector cannot both be mounted on the 8-inch flange if the plane mirror is that far from the sample. We therefore simulated the range of $x$ that the flange allows; because the collection changes little over this range, we fixed $x = 79.6$ mm and varied $\theta$ and $z$. The collection is insensitive to $\theta$ within a few degrees of 39°, whereas it is most sensitive to $z$ and increases as $z$ decreases. We chose $z = 16.5$ mm, the smallest distance at which the PMT cage does not block the light reflected at the edge of the elliptical mirror. In the final geometry, 91.9% of the emitted photons reach the re-entrant window within a spot of about φ20 mm, inside the φ21 mm effective area of the photon counting head (**Fig. 2b**), which leaves a margin of only about 0.5 mm at the edge. This is a geometrical upper limit that excludes mirror reflectivity, window and filter transmittance, and quantum efficiency.

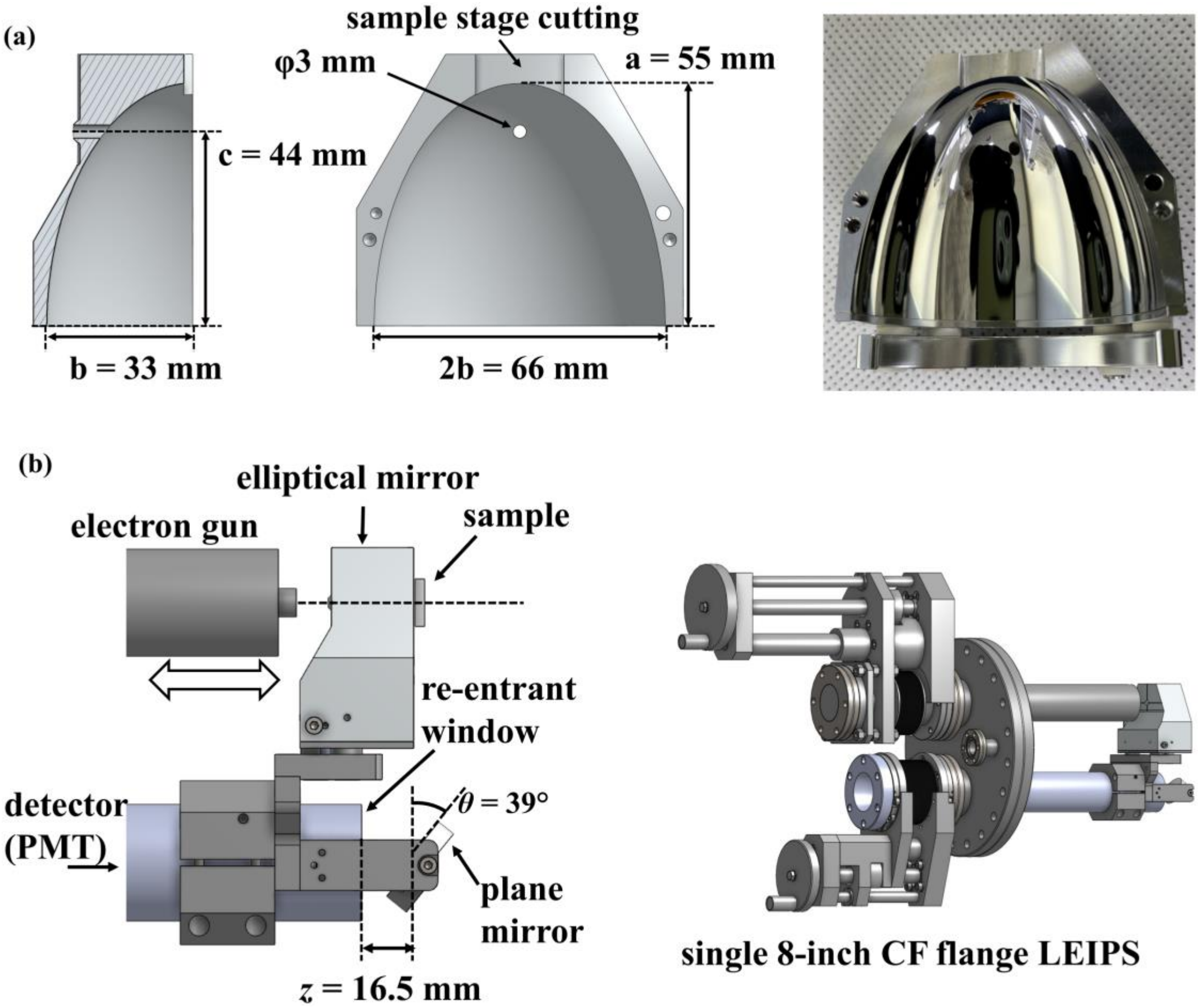

**Figure 3.** Elliptical mirror and single-flange assembly. (a) Side and front views of the elliptical mirror with the semi-major axis *a*, the semi-minor axis *b*, the focal distance *c* and the φ3 mm aperture (left and center), and a photograph of the fabricated mirror (right). (b) Side and isometric views of the assembly on the 8-inch CF flange with the plane mirror tilt *θ* and the distance *z* from the plane mirror to the re-entrant window.

Based on the simulation, the elliptical mirror was fabricated with $a$ = 55 mm, $b$ = 33 mm and $c$ = 44 mm (**Fig. 3a**). The region labeled **"**sample stage cutting**"** is cut away where the mirror would otherwise interfere with the sample stage. This cut allows the stage to bring the sample surface to the first focus while the elliptical mirror still covers as much of the half-space in front of the sample as possible. The plane mirror placed before the second focus folds the ≈130 mm optical path from the sample to the PMT photocathode so that the electron gun and the detector fit on the flange. At the aperture, the curved surface of the mirror is 1 to 2 mm from the sample surface, the smallest clearance that avoids contact. The electron beam from the gun is expected to be about 2 mm in diameter at the sample. The simulated collection is almost the same for aperture diameters of 2 mm and 3 mm, so we chose a 3 mm aperture on the gun axis, slightly larger than the beam. The electrons then pass through the aperture without being absorbed at its wall, as in the OAP mirror [18].

The elliptical mirror was diamond-turned from aluminum, which reflects well over a broad range of wavelengths including the DUV near 250 nm [24]. It was left uncoated because the mirror is biased to guide the electrons, and a dielectric coating would accumulate charge and disturb the low-energy electron beam [18]. The plane KrF mirror has its reflection band centered at 250 nm, so the reflection that folds the path also defines the detection band together with the bandpass filter and rejects stray infrared light from the electron gun filament. Because the light converges toward the second focus, it strikes the plane mirror over a range of incidence angles about the chief ray, which is incident at about 45° because the mirror folds the path by about 90°. The reflection band of a dielectric mirror shifts to shorter wavelength as the angle of incidence increases and differs slightly between s and p polarization, so the rays at the edges of the converging cone see a slightly shifted band, and the effective detection band of the elliptical setup is the product of the filter transmittance and the angle-averaged reflectance of the mirror. This narrows the band slightly at its edges, as the two KrF mirrors do in the OAP-external setup. The ray-tracing assumed 100% reflectance at all angles, so this angle-dependent loss is not included in the predicted collection and acts in the direction of a measured gain smaller than predicted.

After the plane mirror, the light passes through the re-entrant window and the bandpass filter to the photocathode without any lens. The two mirrors are fixed on one support attached to the PMT cage, and the re-entrant window brings the detector to $z = 16.5$ mm from the plane mirror tilted at $\theta = 39°$ (**Fig. 3b**), as determined in the simulation. Once the assembly is mounted in the UHV chamber, the mirrors and their mounts can no longer be adjusted, so they are aligned before installation. The electron gun position, in contrast, needs fine adjustment depending on the voltages applied to the electron gun, the mirror and the sample. The gun is therefore mounted on a linear translator, which moves it in the direction of the double-headed arrow in Fig. 3b. This translation is the only adjustment left after installation, and it allows the gun position to be tuned without breaking the vacuum.

*3.2 Assembly and ex situ optical alignment*

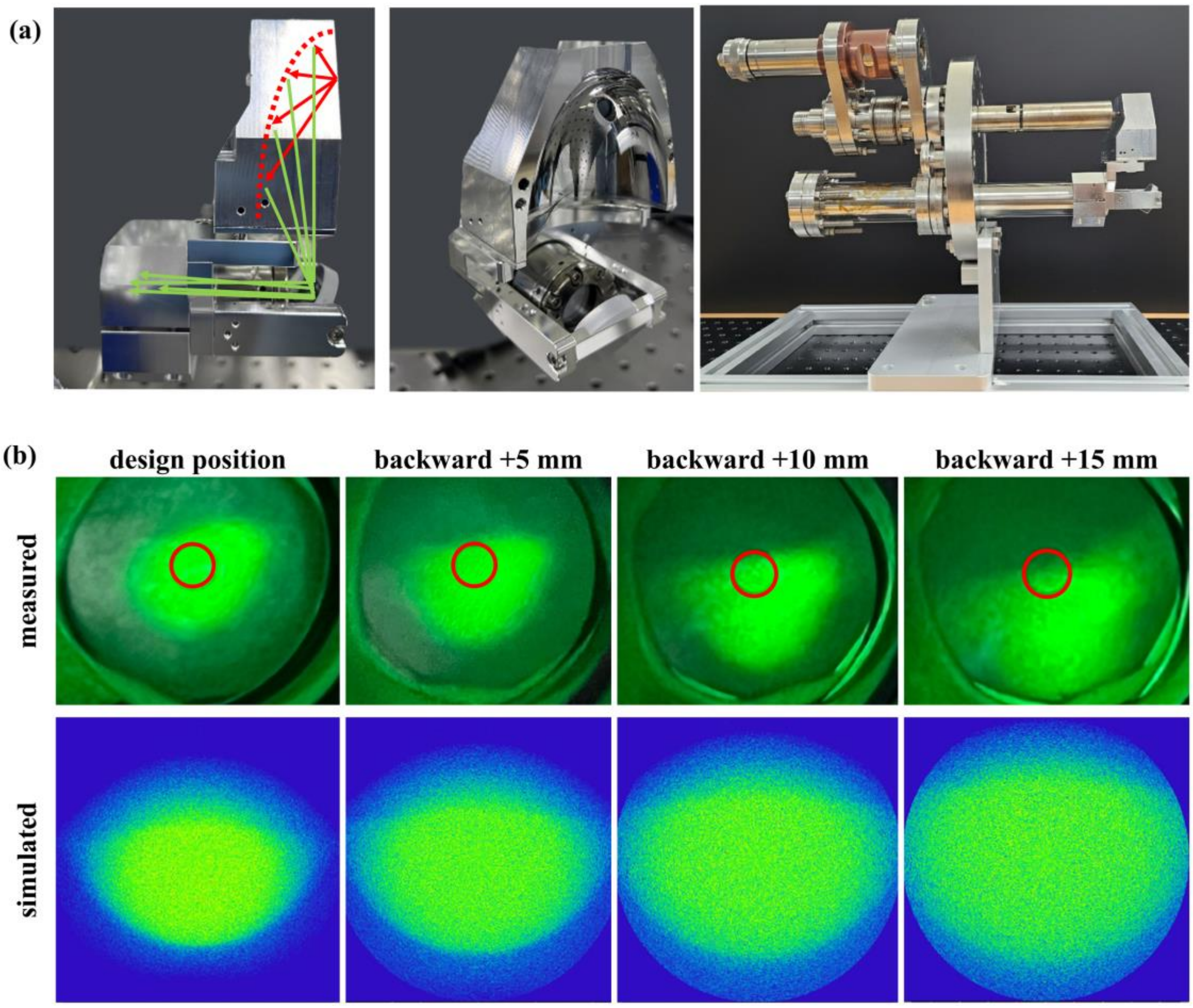

**Figure 4.** Assembly and *ex situ* optical alignment. (a) Side view of the elliptical and plane mirrors on their common support with the optical path drawn over it and the completed assembly. (b) Light spot profiles at the design detector position and 5, 10 and 15 mm behind it. The upper row was measured with a 520 nm laser diode at the electron gun position and diffusely scattering paper at the sample position. The lower row shows the corresponding simulated distributions on axes of ±10.5 mm that span the φ21 mm effective area of the detector. The two rows are not on the same scale. The ≈5 mm red circle on the screen sets the scale of the upper row.

Every optical element must be placed at its design position, because a displacement of a few millimeters along the path lowers the collection efficiency. The mirrors cannot be adjusted in vacuum, so we carefully aligned the optical path on the bench before installation (**Fig. 4a**). A 520 nm laser diode at the position of the electron gun illuminated diffusely scattering paper at the position of the sample, which approximates the Lambertian emission from the sample. We adjusted the laser spot on the paper to correspond as closely as possible to the 2 mm source assumed in the ray-tracing. The KrF plane mirror reflects well only in a narrow band near 250 nm, so we replaced it with a UV-enhanced Al mirror that reflects over a broad range of wavelengths including 520 nm. The angle of reflection does not depend on wavelength, so the path is the same as for the detection band. **Fig. 4b** shows the measured and simulated light spot on a screen at the detector position. The spot broadens monotonically as the screen moves 5, 10 and 15 mm behind the design position. This agrees with the ray-tracing and shows that the PMT also requires careful positioning, because the spot nearly fills the effective area of the detector. The position of the sample matters even more in daily use because the sample is exchanged for every measurement. A displacement of the sample along the surface normal is limited toward the mirror by the 1–2 mm clearance to the mirror aperture, and a displacement away from the mirror moves the source off the first focus and defocuses the image at the detector in the same way as the detector displacement in Fig. 4b. We did not simulate the collection as a function of the sample displacement; in practice the sample is returned to the same manipulator position for every measurement. We then installed the assembly in the chamber with the mirrors and the PMT at their design positions and measured the LEIPS spectra of $C_{60}$ films.

*3.3 LUMO intensities of the four setups*

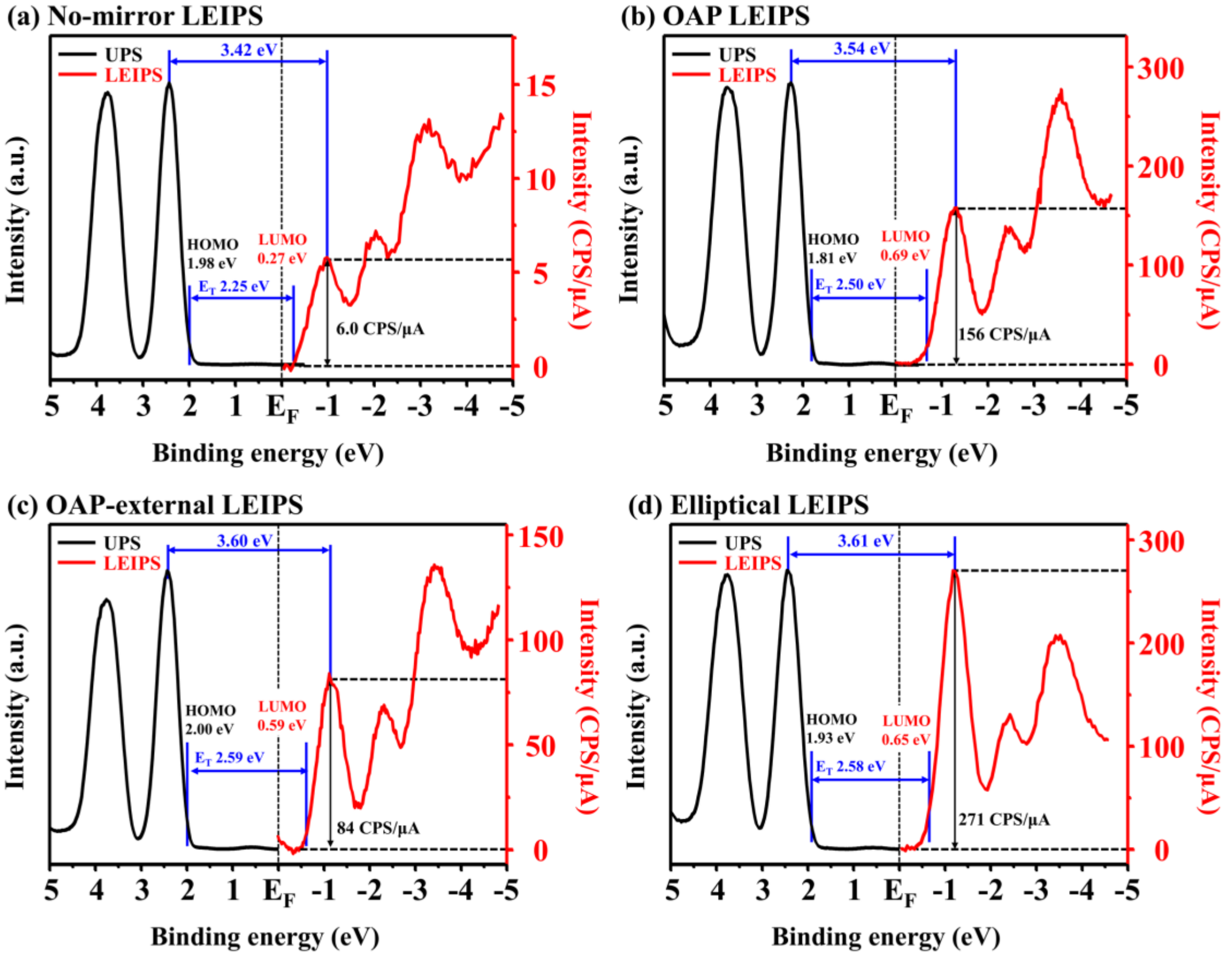

**Figure 5.** UPS and LEIPS spectra of $C_{60}$ films on Ag. (a) No-mirror, (b) OAP, (c) OAP-external and (d) elliptical setups. A constant background has been subtracted from each LEIPS spectrum. The right vertical axis is the LEIPS count rate per μA of sample current (CPS/μA), and its scale differs between panels. The black arrow marks the LUMO peak height.

**Fig. 5** shows the UPS and LEIPS spectra of $C_{60}$ films on Ag measured with the four setups. The LUMO peak intensity is 6.0, 156, 84 and 271 CPS/μA for the no-mirror, OAP, OAP-external and elliptical setups. The elliptical setup therefore gives 45 times the intensity of the no-mirror setup, 1.74 times that of the OAP setup and 3.23 times that of the OAP-external setup. Table 1 summarizes the optical elements, detectors and acquisition parameters of the four setups together with these intensities. From the raw count rates at the LUMO peak (about 130 and 270 counts per second) and the total dwell time per energy step (40 and 20 s), the counting-statistics uncertainty of the peak height is about 1.5% for the elliptical and OAP setups, so the statistical uncertainty of the ratio 1.74 is only about 2%. The variation between films and between alignments is larger but could not be quantified, because each setup was measured on a single film; the ratios should therefore be read as representative values rather than as precision measurements.

**Table 1.** Optical elements, detectors and acquisition parameters of the four setups, with the measured LUMO intensities and the intensities expected relative to the OAP setup.

| | **No-mirror** | **OAP** | **OAP-external** | **Elliptical** |
|---|---|---|---|---|
| Collection optics | None | OAP mirror | OAP mirror, two KrF mirrors and two lenses in air | Elliptical mirror and one KrF plane mirror |
| Geometric collection (%)[a] | 3.06 | 63.3 | 63.3[b] | 91.9 |
| Bandpass filter (peak transmittance, FWHM) | BP-250 (58%, 9.4 nm) | BP-250 (58%, 9.4 nm) | BP-250 (58%, 9.4 nm) | Hg01-254 (78%, 15.7 nm) |
| Detector | R6834 head-on PMT | R6834 head-on PMT | R7154P side-on PMT | H11870-09 photon counting head |
| **Quantum efficiency at 254 nm (%)**[c] | ≈14 (Cs-Te) | ≈14 (Cs-Te) | 30 (Cs-Te) | ≈14 (Cs-Te) |
| **Dark count rate ($s^{-1}$)**[c] | n.s. (0.3 nA anode dark current) | n.s. (0.3 nA anode dark current) | 5 (typ.), 20 (max.) | 15 (typ.) |
| Sample current (μA) | 2.97 | 1.75 | 1.55 | 0.49 |
| Energy step / dwell time / sweeps | 0.05 eV / 1 s / 160 | 0.025 eV / 1 s / 20 | 0.025 eV / 1 s / 20 | 0.05 eV / 0.2 s / 200 |
| LUMO intensity (CPS/μA) | 6.0 | 156 | 84 | 271 |
| Measured intensity relative to OAP | 0.038 | 1 | 0.54 | 1.74 |
| Expected intensity relative to OAP (geometry × filter peak transmittance × quantum efficiency) | 0.048 | 1 | ≈2.1[b] | ≈1.95 |

[a] Fraction of the emitted photons reaching the detector window in the ray-tracing. [b] Light leaving the chamber; the external path to the side-on PMT was not simulated. [c] From the manufacturer's datasheets. The quantum efficiency of the R6834 was converted from its cathode radiant sensitivity at 254 nm (28 mA/W); the H11870-09 contains a Cs-Te photocathode with a similar spectral response (185–320 nm, peak at 240 nm), for which the datasheet gives the count-sensitivity curve rather than the quantum efficiency, so the same value is assumed. n.s., not specified for photon counting; the typical anode dark current is given instead. The H11870-09-Y003 is a custom variant of the H11870-09.

The four spectra were measured on different days with different detectors. They can nevertheless be compared because the highest occupied molecular orbital (HOMO)–LUMO peak separation, which does not depend on where an onset is placed, is 3.54, 3.60 and 3.61 eV for the OAP, OAP-external and elliptical setups, and their transport gaps (2.50, 2.59 and 2.58 eV) agree within 0.09 eV. The no-mirror setup stands apart, with a peak separation of 3.42 eV and a transport gap of 2.25 eV. Its film probably degraded because it received 2.97 μA for about 6 h, a much higher current and longer time than the others, so the factor of 45 is probably an overestimate. Because the elliptical setup gives the highest LUMO intensity, its beam current could be reduced to 0.49 μA over the 93 min acquisition, during which the OAP setup received 1.75 μA; the charge delivered in the elliptical setup is thus 28% of that in the OAP setup. The transport gaps of the two setups agree, showing no resolvable difference in degradation.

Ray-tracing predicts that the elliptical setup delivers 1.45 times as many photons to the detector as the OAP setup (91.9% versus 63.3%), and the measured ratio is 1.74. The two setups differ, however, in more than the collection optics (Table 1): the bandpass filter of the elliptical setup has a higher peak transmittance (78% versus 58%) and a broader passband (15.7 versus 9.4 nm,

Fig. S1), and the detector is a different model. The product of the geometric ratio and the ratio of the peak transmittances, $1.45 \times 1.34 = 1.95$, already exceeds the measured 1.74, and the broader passband should raise the count rate further, because at each electron energy the count rate scales roughly with the transmittance integrated over the passband. The measured gain is therefore smaller than the geometry and the filter together predict. The detector is unlikely to explain much of this difference, because the R6834 and the H11870-09 both use a Cs-Te photocathode with a similar spectral response (peak at 240 nm) and a comparable quantum efficiency at 254 nm (Table 1), although the counting efficiency set by the fixed discriminator level of the photon counting head was not measured. The variation between films and alignments noted above may account for part of it. The angle-dependent reflectance of the KrF plane mirror at the edges of the converging cone, taken as 100% in the ray-tracing, acts in the same direction. Because the filter and the detector were changed together with the mirror, the present data cannot separate the contribution of the collection geometry from that of the other components. What the comparison does establish is that the elliptical setup as built gives the highest LUMO intensity of the four, and that the measured gain is of the same order as the geometric gain predicted by ray-tracing. For the OAP-external setup, the collection of the OAP mirror and the twofold quantum efficiency of the side-on PMT predict about twice the signal of the OAP setup, but the measured signal is only 0.54 times. The two KrF mirrors of the OAP-external setup narrow the passband slightly, but their reflectivity above 99% keeps their loss small. The discrepancy between calculation and measurement is thus far larger for the long external path of the OAP-external setup than for the short in-vacuum path of the elliptical setup. A likely contribution is residual divergence of the beam leaving the OAP mirror: the calculation treats this beam as collimated, but the 2 mm emitting spot and the alignment tolerance of the mirror broaden it as the path length increases, so part of the beam may fall outside the clear apertures of the lenses and mirrors. A more specific loss is the small photocathode of the side-

on PMT: the effective area of the R7154P is a rectangle of only 8 × 24 mm (Table 1), and the two lenses must refocus an extended, divergent beam onto it, so part of the refocused image probably falls outside the photocathode. Overfilling of the photocathode, rather than divergence in general, may therefore be the dominant loss. A ray-tracing of the external optics with the same 2 mm source, or a bench measurement of its throughput with a lamp, would settle this; neither has been done, so this explanation remains a hypothesis. These losses, none of which the calculation includes, more than offset the twofold quantum efficiency of the side-on PMT. The elliptical setup gives the highest sensitivity so far because it combines the largest collection with a short optical path kept inside the vacuum, over which the divergence of the beam has little room to grow.

The energy resolution of LEIPS is determined by the energy spread of the electron beam together with the detection bandwidth, which is set mainly by the bandpass filter and the KrF plane mirror; if both contributions are approximately Gaussian, the overall resolution is their quadrature sum, and the width of the Fermi edge of a clean metal, which is intrinsically only a few $kT$ wide, measures this overall resolution directly. The optical bandwidth alone is ≈190 meV for the BP-250 filter and ≈300 meV for the Hg01-254 filter. Compared with the OAP-external setup, the elliptical setup uses one KrF plane mirror instead of two and a bandpass filter with a broader passband (Fig. S1), and both changes broaden the detection band. The FWHM of the $C_{60}$ LUMO, from Gaussian fits after background subtraction, is 0.74, 0.67, 0.61 and 0.67 eV for the no-mirror, OAP, OAP-external and elliptical setups (Fig. S2): the value for the elliptical setup equals that of the OAP setup, and its 0.06 eV excess over the OAP-external setup is no larger than the difference between the OAP and OAP-external setups, which used the same filter. The Fermi edge of clean Ag, a much sharper feature than the $C_{60}$ LUMO, shows the same trend: the FWHM of a Gaussian fitted to its derivative dI/dE is 0.409, 0.384, 0.355 and 0.396 eV for the four setups (Fig. S3), and the value for the elliptical setup lies within the range of the three setups with the narrower filter. Subtracting the optical bandwidth in quadrature and neglecting the thermal width of the edge gives an electron energy spread of about 0.30–0.36 eV for the three setups with the BP-250 filter and about 0.26 eV for the elliptical setup. The energy spread of the electron beam is therefore the larger contribution to the resolution, and the lower sample current of the elliptical setup (0.49 μA against 1.75 μA for the OAP setup), which reduces the space-charge broadening of the beam, may have offset its broader detection band, so that the overall resolution was unchanged.

For comparison with other LEIPS instruments, the bandpass-filter apparatus of Yoshida [17] has an overall resolution of 0.27 eV and the angle-resolved apparatus of Kashimoto et al. [19], which also collects the light with an elliptical mirror, 0.23 eV, whereas the corresponding Fermi-edge width of the present instrument is 0.396 eV (Fig. S3). A comparison of count rates is only indicative, because the detection bandwidth, the electron current, the collection solid angle and the film preparation differ between laboratories; the present value of 271 CPS/μA for a 10 nm $C_{60}$ film at 0.49 μA is therefore given as a reference point for other groups rather than as a ranking. As for the footprint, the electron gun, both mirrors and the photon counting head of the elliptical setup are carried by a single 8-inch CF flange, the optical path from the sample to the detector is 130 mm, and the only parts inside the chamber are the gun and the mirror assembly (Fig. 3); the OAP-external setup, by contrast, needed two lenses and two mirrors in air between the vacuum window and the side-on PMT, all of which had to be aligned on the air side (Fig. 1c).

## 4. **Conclusion**

We developed a compact LEIPS instrument in which an elliptical mirror and a plane folding mirror deliver the photons from the sample to the detector, and the electron gun, collection optics and detector are mounted on a single 8-inch flange. Four photon-collection configurations were compared using the LUMO peak height of $C_{60}$ films thermally evaporated under UHV, with all steps from deposition to measurement performed *in situ*. Our previous OAP mirror delivered the emitted photons to the detector as a collimated beam and raised the fraction reaching the detector from 3.06% without a mirror to 63.3%. An intermediate configuration with external optics, introduced to use a side-on PMT with about twice the quantum efficiency, reduced the signal to 0.54 times that of the OAP configuration, most likely because the extended, divergent beam from the OAP mirror could not be refocused completely onto the small photocathode of the side-on PMT and lost light at each element of its long external path. We therefore adopted an in-vacuum elliptical mirror that focuses the photons onto the detector over a short path without lenses, which raised the geometric collection to 91.9%. A KrF plane mirror with high reflectance in the 250 nm band folds the converging light by about 90°, so that the detector lies parallel to the electron gun on the same flange, and the geometry was optimized by ray-tracing with Zemax OpticStudio within the mechanical constraints of the flange. The elliptical configuration gave a LUMO intensity of 271 CPS/μA, 1.74 times that of the OAP configuration and about 45 times that of the no-mirror configuration, the highest among the configurations tested so far, without a resolvable loss of energy resolution. Because the filter and the detector were changed together with the mirror, this gain cannot be attributed to the collection geometry alone, but it is of the same order as the 1.45-fold gain in collected photons predicted by ray-tracing. These results show that high-sensitivity

LEIPS requires both a large collection solid angle and a short optical path kept inside the vacuum. Because the detector remains in air directly behind the re-entrant window, it can be replaced with one of higher quantum efficiency without modifying the in-vacuum optics.

**CRediT authorship contribution statement**

Jong-Am Hong: Conceptualization, Methodology, Investigation, Writing – original draft. Min-Jae Maeng: Validation, Investigation, Writing – original draft, Writing – review & editing. Kyu-Myung Lee: Data curation, Formal analysis. Seunggu Lee: Data curation, Formal analysis. In-Sung Park: Data curation, Formal analysis. Yongsup Park: Conceptualization, Supervision, Project administration, Funding acquisition, Writing – review & editing.

**Declaration of competing interest**

The authors declare the following financial interests/personal relationships which may be considered as potential competing interests: Yongsup Park has patent #KR 10-2023-0179334 pending to University-Industry Cooperation Group of Kyung Hee University. Co-inventor on the pending patent application KR 10-2023-0179334 (assignee: University-Industry Cooperation Group of Kyung Hee University) - Jong-Am Hong If there are other authors, they declare that they have no known competing financial interests or personal relationships that could have appeared to influence the work reported in this paper.

**Declaration of generative AI and AI-assisted technologies in the writing process**

During the preparation of this work the authors used Claude (Anthropic) in order to edit the language of the manuscript and check its internal consistency. After using this tool, the authors reviewed and edited the content as needed and take full responsibility for the content of the publication.

**Acknowledgements**

This work was supported by the Korea Basic Science Institute (National Research Facilities and Equipment Center) grant funded by the Ministry of Education (2021R1A6C101A437).

**Appendix A. Supplementary data**

Supplementary data to this article can be found online.

**Data availability**

Data will be made available on request.

**References**

[1] G. D'Avino, L. Muccioli, F. Castet, C. Poelking, D. Andrienko, Z.G. Soos, J. Cornil, D. Beljonne, Electrostatic phenomena in organic semiconductors: fundamentals and implications for photovoltaics, J. Phys.: Condens. Matter 28 (2016) 433002. https://doi.org/10.1088/0953-8984/28/43/433002.

[2] Y. Gao, Surface analytical studies of interfaces in organic semiconductor devices, Mater. Sci. Eng. R Rep. 68 (2010) 39–87. https://doi.org/10.1016/j.mser.2010.01.001.

[3] J.L. Brédas, J. Cornil, A.J. Heeger, The exciton binding energy in luminescent conjugated polymers, Adv. Mater. 8 (1996) 447–452. https://doi.org/10.1002/adma.19960080517.

[4] V. Dose, VUV isochromat spectroscopy, Appl. Phys. 14 (1977) 117–118. https://doi.org/10.1007/BF00882639.

[5] N.V. Smith, Inverse photoemission, Rep. Prog. Phys. 51 (1988) 1227–1294. https://doi.org/10.1088/0034-4885/51/9/003.

[6] K. Yamada, S. Yanagisawa, T. Koganezawa, K. Mase, N. Sato, H. Yoshida, Impact of the molecular quadrupole moment on ionization energy and electron affinity of organic thin films: Experimental determination of electrostatic potential and electronic polarization energies, Phys. Rev. B 97 (2018) 245206. https://doi.org/10.1103/PhysRevB.97.245206.

[7] J.B. Pendry, Theory of inverse photoemission, J. Phys. C: Solid State Phys. 14 (1981) 1381–1391. https://doi.org/10.1088/0022-3719/14/9/022.

[8] K. Desinger, V. Dose, M. Glöbl, H. Scheidt, Momentum resolved bremsstrahlung isochromat spectra from Ni(0 0 1), Solid State Commun. 49 (1984) 479–481. https://doi.org/10.1016/0038-1098(84)90668-9.

[9] J.A. Lipton-Duffin, A.G. Mark, G.K. Mullins, G.E. Contant, A.B. McLean, An inverse photoemission system with large solid angle of detection and adjustable optical bandpass, Rev. Sci. Instrum. 75 (2004) 445–454. https://doi.org/10.1063/1.1637435.

[10] H. Yoshida, Near-ultraviolet inverse photoemission spectroscopy using ultra-low energy electrons, Chem. Phys. Lett. 539–540 (2012) 180–185. https://doi.org/10.1016/j.cplett.2012.04.058.

[11] H. Yoshida, Principle and application of low energy inverse photoemission spectroscopy: A new method for measuring unoccupied states of organic semiconductors, J. Electron Spectrosc. Relat. Phenom. 204 (2015) 116–124. https://doi.org/10.1016/j.elspec.2015.07.003.

[12] H. Yoshida, Measuring the electron affinity of organic solids: An indispensable new

tool for organic electronics, Anal. Bioanal. Chem. 406 (2014) 2231–2237. https://doi.org/10.1007/s00216-014-7659-1.

[13] C. Thiede, A.B. Schmidt, M. Donath, Optimizing the performance of bandpass photon detectors for inverse photoemission: Transmission of alkaline earth fluoride window crystals, Rev. Sci. Instrum. 86 (2015) 085101. https://doi.org/10.1063/1.4927459.

[14] C. Thiede, I. Niehues, A.B. Schmidt, M. Donath, The acetone bandpass detector for inverse photoemission: operation in proportional and Geiger–Müller modes, Meas. Sci. Technol. 29 (2018) 065901. https://doi.org/10.1088/1361-6501/aab941.

[15] I.G. Hill, A.B. McLean, A comparison of two high performance inverse photoemission bandpass detectors, Rev. Sci. Instrum. 69 (1998) 261–264. https://doi.org/10.1063/1.1148506.

[16] H. Yoshida, Low-energy inverse photoemission spectroscopy using a high-resolution grating spectrometer in the near ultraviolet range, Rev. Sci. Instrum. 84 (2013) 103901. https://doi.org/10.1063/1.4822119.

[17] H. Yoshida, Note: Low energy inverse photoemission spectroscopy apparatus, Rev. Sci. Instrum. 85 (2014) 016101. https://doi.org/10.1063/1.4860055.

[18] J.-A. Hong, K.-M. Lee, M.-J. Maeng, Y. Park, Enhanced sensitivity in low-energy inverse photoemission spectroscopy with an off-axis parabolic mirror for efficient light collection, Adv. Mater. Technol. 10 (2025) 2402143. https://doi.org/10.1002/admt.202402143.

[19] Y. Kashimoto, S. Ideta, H. Sato, H. Orio, K. Kawamura, H. Yoshida, High-energy-resolution angle-resolved inverse-photoelectron spectroscopy apparatus for damage-free measurements of conduction band structures of functional materials, Rev. Sci.

Instrum. 94 (2023) 063903. https://doi.org/10.1063/5.0138204.

[20] H. Sato, S.A. Abd. Rahman, Y. Yamada, H. Ishii, H. Yoshida, Conduction band structure of high-mobility organic semiconductors and partially dressed polaron formation, Nat. Mater. 21 (2022) 910–916. https://doi.org/10.1038/s41563-022-01308-z.

[21] A. Zumbülte, A.B. Schmidt, M. Donath, Momentum resolution in inverse photoemission, Rev. Sci. Instrum. 86 (2015) 013908. https://doi.org/10.1063/1.4906508.

[22] N. Babbe, W. Drube, I. Schäfer, M. Skibowski, A simple and compact system for combined angular resolved inverse photoemission and photoemission in the vacuum ultraviolet, J. Phys. E: Sci. Instrum. 18 (1985) 158. https://doi.org/10.1088/0022-3735/18/2/014.

[23] H. Yoshida, Low-energy inverse photoemission study on the electron affinities of fullerene derivatives for organic photovoltaic cells, J. Phys. Chem. C 118 (2014) 24377–24382. https://doi.org/10.1021/jp509141y.

[24] H. Ehrenreich, H.R. Philipp, B. Segall, Optical properties of aluminum, Phys. Rev. 132 (1963) 1918. https://doi.org/10.1103/PhysRev.132.1918.